\documentclass{pasj01}

\Received{$\langle$reception date$\rangle$}
\Accepted{$\langle$acception date$\rangle$}
\Published{$\langle$publication date$\rangle$}

\usepackage{natbib,bm}
\usepackage[varg]{txfonts}  

\usepackage[pdftex]{graphicx,color}  

\definecolor{pinegreen}{cmyk}{0.92, 0.0, 0.59,0.25}

\newcommand{\pd}{\partial}
\newcommand{\au}{{\rm au}} 
\newcommand{\kB}{k_{\rm B}}
\newcommand{\paren}[1]{\left({#1}\right)} 
\newcommand{\subsc}[2]{#1_{{\rm #2}}}
\newcommand{\eqref}[1]{(\ref{#1})}

\begin{document}

\title{A self-consistent model for dust settling and the vertical shear instability in protoplanetary disks}

\author{Yuya \textsc{Fukuhara}\altaffilmark{1}}
\author{Satoshi \textsc{Okuzumi}\altaffilmark{1}}

\altaffiltext{1}{Department of Earth and Planetary Sciences, Tokyo Institute of Technology, Meguro, Tokyo 152-8551, Japan}
\email{fukuhara.y.ab@m.titech.ac.jp}


\maketitle

\begin{abstract}
    The spatial distribution of dust particles in protoplanetary disks affects dust evolution and planetesimal formation processes. 
    The vertical shear instability (VSI) is one of the candidate hydrodynamic mechanisms that can generate turbulence in the outer disk region and affect dust diffusion. 
    Turbulence driven by the VSI has a predominant vertical motion that can prevent dust settling. 
    On the other hand, the dust distribution controls the spatial distribution of the gas cooling rate, thereby affecting the strength of VSI-driven turbulence. 
    Here, we present a semi-analytic model that determines the vertical dust distribution and the strength of VSI-driven turbulence in a self-consistent manner. 
    The model uses an empirical formula for the vertical diffusion coefficient in VSI-driven turbulence obtained from our recent hydrodynamical simulations. 
    The formula returns the vertical diffusion coefficient as a function of the vertical profile of the cooling rate, which is determined by the vertical dust distribution. 
    We use this model to search for an equilibrium vertical dust profile where settling balances with turbulent diffusion for a given maximum grain size.  
    We find that if the grains are sufficiently small, there exists a stable equilibrium dust distribution where VSI-driven turbulence is sustained at a level of $\alpha_z \sim 10^{-3}$, where $\alpha_z$ is the dimensionless vertical diffusion coefficient. 
    However, as the maximum grain size increases, the equilibrium solution vanishes because the VSI can no longer stop the settling of the grains. 
    This runaway settling may explain highly settled dust rings found in the outer part of some protoplanetary disks.
\end{abstract}


\section{Introduction}\label{sec:intro}
The initial stage of planet formation is the formation of kilometer-sized planetesimals from micron-sized dust grains (for reviews, \citealt{Johansen+2014,DrazkowskaBitsch+:2023aa}).
These dust growth and planetesimal formation depend on the vertical distribution of dust particles.
If the dust density at the midplane well exceeds the gas density, the streaming and gravitational instabilities set in (e.g., \citealt{Goldreich:1973aa,Sekiya:1998aa,Youdin:2002aa,YoudinGoodman:2005aa}), leading to planetesimal formation.
Understanding what processes determine the dust vertical profile in disks is also essential for interpreting recent high-resolution radio observations showing that the dust at different radial locations or in different disks is settled to different extents (\citealt{Pinte:2016aa,DoiKataoka:2021oz,VillenaveStapelfeldt+:2022pp,VillenavePodio+:2023aa,PizzatiRosotti+:2023aa}, and for a review, \citealt{MiotelloKamp+:2023aa}).

In protoplanetary disks, the vertical profile of dust particles critically depends on the intensity of gas disk turbulence causing dust diffusion.
The question then is what mechanisms generate disk turbulence.
Recent theoretical studies have shown that not only the magnetorotational instability (MRI; \citealt{BalbusHawley1991}) but also some thermo-hydrodynamical instabilities are important turbulence-driving mechanisms (for reviews, \citealt{LyraUmurhan2019,LesurFlock+:2023aa}).
Among them, the vertical shear instability (VSI; \citealt{UrpinBrandenburg1998,ArltUrpin2004,NelsonGresselUmurhan2013,LinYoudin2015}) is thought to be the leading mechanism for driving turbulence in outer disk regions where the current radio observations best constrain the degree of dust settling.
The VSI requires rapid cooling of disk gas in addition to a vertically varying gas orbital velocity \citep{Urpin2003,NelsonGresselUmurhan2013,LinYoudin2015,MangerPfeil+:2021cm}.
Therefore, the VSI tends to operate in the outer disk region with low optical depths \citep{Malygin+2017,PfeilKlahr2019,FukuharaOkuzumi+:2021ca,Melon-FuksmanFlock+:2024ab}, where it may dominate over the MRI (e.g., \citealt{CuiBai:2022aa}).
The VSI generates turbulence with a predominant vertical gas motion (e.g., \citealt{NelsonGresselUmurhan2013,StollKley2014}) that can prevent dust settling \citep{StollKley:2016vp,FlockNelson+2017,Flock:2020aa,DullemondZiampras+:2022aa}.
 This turbulence can also stir dust particles significantly \citep{StollKley:2016vp,FlockNelson+2017} and thus induce their collisional velocities, leading to suppression of planetesimal formation through coagulation (e.g, \citealt{OrmelCuzzi2007,Brauer:2008aa,Okuzumi:2012aa}).
On the other hand, VSI-driven turbulence can produce both small short-lived and azimuthally large long-lived vortices \citep{Richard:2016aa,MangerKlahr:2018dw,Flock:2020aa,PfeilKlahr:2021nr,Melon-FuksmanFlock+:2024aa}.
These vortices may lead to dust concentration and subsequent planetesimal formation through gravitational collapse (e.g., \citealt{BargeSommeria:1995qd,RaettigLyra+:2021sb,LehmannLin:2022nr}).

Importantly, dust particles are the dominant opacity source and determine the local cooling rates of protoplanetary disks \citep{Malygin+2017,BarrancoPei+:2018kc}. 
Therefore, their spatial distribution controls where in the disks the VSI operates \citep{PfeilKlahr2019,FukuharaOkuzumi+:2021ca} and can even affect the dust vertical diffusivity at the midplane  \citep{PfeilKlahr:2021nr,FukuharaOkuzumi+:2023aa,PfeilBirnstiel+:2023aa}.
The consequence of this dust--VSI thermal interaction for dust settling remains unclear.

In this paper, we model the above-mentioned interaction between dust and the VSI to study how dust setting and diffusion balance in VSI-driven turbulence.
To this end, we present a semi-analytic model that determines the vertical dust distribution and the strength of VSI-driven turbulence in a self-consistent manner. 
The model uses an empirical formula for the vertical diffusion coefficient in VSI-driven turbulence obtained from our recent hydrodynamical simulations \citep{FukuharaOkuzumi+:2023aa}. 
The formula returns the vertical diffusion coefficient as a function of the vertical profile of the cooling rate, which is determined by the vertical dust distribution. 
We use this model to search for an equilibrium vertical dust profile where settling balances with turbulent diffusion for a given grain size.

This paper is organized as follows. 
In section \ref{sec:method}, we describe our self-consistent model. 
We present the main results in section \ref{sec:results} and discuss the implication of our study in section \ref{sec:discussion}.
Section \ref{sec:summary} presents a summary.

\section{Method}\label{sec:method}

\begin{figure*}[t]
    \begin{center}
    \includegraphics[width=\hsize,bb = 0 0 1405 853]{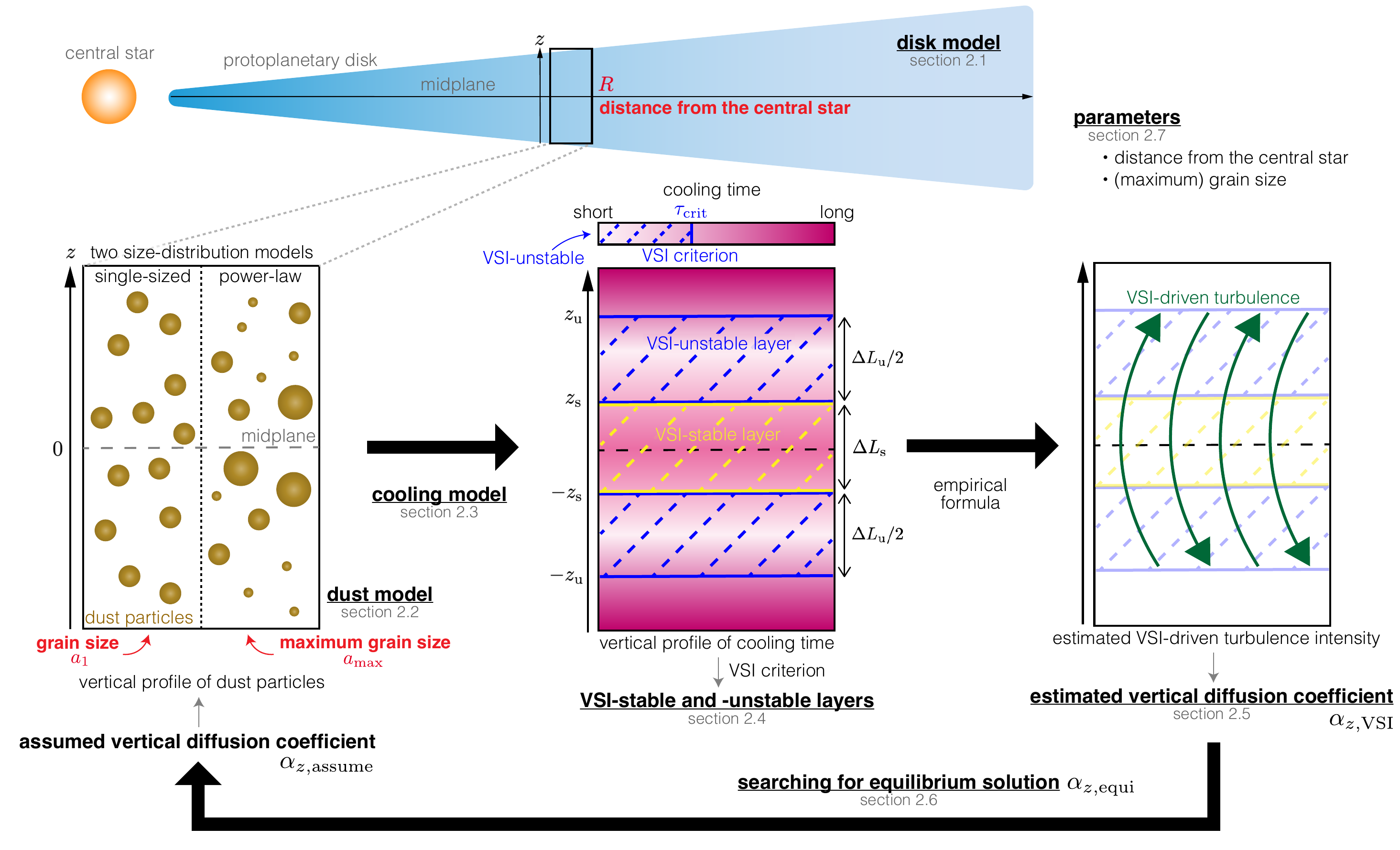}
    \end{center}
    \caption{Overview showing the self-consistent model determining the vertical dust profile and VSI-driven turbulence intensity in this study. The model considers a protoplanetary disk consisting of gas and dust (section \ref{subsec:disk_model}). Assuming dust particle size and vertical diffusion coefficient $\alpha_{z,\rm assume}$, we compute the vertical profile of dust particles for the single-sized model and power-law size distribution model (section \ref{subsec:dust_model}). The dust vertical profile determines the vertical profile of the cooling timescale (section \ref{subsec:cooling_model}) and thereby localizes the region where the VSI operates (VSI-unstable layer), using the linear VSI criterion (section \ref{subsec:unstable_region}). Based on the empirical formula derived from the hydrodynamical simulations, we estimate the vertical diffusion coefficient of VSI-driven turbulence $\alpha_{z,\rm VSI}$ (section \ref{subsec:dust_vertical_diffusion}). Comparing estimated coefficient $\alpha_{z,\rm VSI}$ with assumed coefficient $\alpha_{z,\rm assume}$, we search an equilibrium vertical dust profile where settling balances with turbulent diffusion for a given grain size (section \ref{subsec:Calculation_procedure}). Our parameters are the grain size for the single-sized model or the maximum grain size for the power-law size distribution model, and the distance from the central star (section \ref{subsec:computational_domain and_parameter_choices}).}
    \label{fig:fig1_Schematic_view}
\end{figure*}

In this section, we describe our self-consistent model that determines the dust vertical distribution and VSI-driven turbulence intensity simultaneously (see figure \ref{fig:fig1_Schematic_view} for an overview).
We assume a protoplanetary disk consisting of gas (section \ref{subsec:disk_model}) and dust (section \ref{subsec:dust_model}).
The dust grains dominate the disk's opacity (section \ref{subsec:dust_model}) and their distribution determines the disk's local cooling rate distribution (section \ref{subsec:cooling_model}).
Because the VSI requires rapid gas cooling, the cooling rate distribution in turn determines the location where the linear VSI operates (section \ref{subsec:unstable_region}), which we call the VSI-unstable layer\footnote{\citet{FukuharaOkuzumi+:2021ca} called these the VSI zones.}.
The VSI produces turbulent gas motion propagating across the boundary between the VSI-unstable and stable regions. 
We calculate the VSI-driven vertical diffusion coefficient from the thicknesses of the VSI-unstable and -stable layers using an empirical formula based on our previous hydrodynamical simulations (section \ref{subsec:dust_vertical_diffusion}).
The vertical diffusion coefficients can be used to update the vertical dust distribution. 
Iterating these calculating steps, we determine the vertical dust distribution and the strength of VSI-driven turbulence in a self-consistent manner (section \ref{subsec:Calculation_procedure}).
In the following subsections, we describe the model in more detail.

    \subsection{Disk model}\label{subsec:disk_model}
    We consider an axisymmetric disk around a solar-mass star.
    We adopt the cylindrical coordinate system $(R,~z)$, where $R$ and $z$ are the distance from the central star and the height from the midplane, respectively.
    The gas surface density is given by
    \begin{equation}\label{eq:Sigmagas}
        \Sigma_{\rm g}(R) =  \frac{\paren{2-\beta_{\Sigma}}\subsc{M}{disk}}{2\pi R_c^2}\paren{\frac{R}{R_c}}^{-\beta_{\Sigma}} \exp{\left[-\paren{\frac{R}{R_c}}^{2-\beta_{\Sigma}}\right]} ,
    \end{equation}
    where $\subsc{M}{disk}$ is the total mass of the gas disk, $R_c$ is the characteristic radius, and $\beta_\Sigma$ is a dimensionless number characterizing the radial slope of the gas surface density. 
    We fix the gas disk parameters to $\subsc{M}{disk} = 0.01\MO$, $R_c = 100~ \au$, and $\beta_\Sigma = 1$.
    
    Assuming that the disk is optically thick to stellar radiation and that the stellar luminosity is equal to the solar luminosity, the temperature of the disk is given by
    \begin{equation}\label{eq:Tgas}
        T(R) = T_0\paren{\frac{R}{1{\rm ~au}}}^q,
    \end{equation}
    with $T_0 = 130 {\rm ~K}$ and $q = -3/7$  \citep{Chiang:1997aa}. 
    We assume that the disk is vertically isothermal, ignoring warmer surface layers that are optically thin to the starlight \citep{Chiang:1997aa}. 
    
    From vertical hydrostatic equilibrium, the gas density is given by
    \begin{equation}\label{eq:rhogas}
        \rho_{\rm g}(R,~z) = \frac{\Sigma_{\rm g}}{\sqrt{2\pi}H_{\rm g}} \exp{\left(-\frac{z^2}{2H_{\rm g}^2}\right)}.
    \end{equation}
    where $H_{\rm g} = c_{\rm s}/\Omega_{\rm K}$ is the gas scale height with $c_{\rm s}$ and $\Omega_{\rm K}$ being the isothermal sound speed and Keplerian frequency, respectively.
    The isothermal sound speed is given by $c_{\rm s} = \sqrt{k_{\rm B}T/m_{\rm g}}$, where $k_{\rm B}$ is the Boltzmann constant and $m_{\rm g}$ is the mean molecular mass of the gas.
    The Keplerian frequency is given by $\Omega_{\rm K} = \sqrt{GM_*/R^3}$, where $G$ is the gravitational constant and $M_*$ is the mass of the central star.
    In this study, $m_{\rm g}$ and $M_*$ are taken to be $2.3m_{\rm p}$ and $1\MO$, respectively, where $m_{\rm p}$ is the proton mass.

    \subsection{Dust model}\label{subsec:dust_model}
    We here describe the dust model that we use to calculate the cooling timescale (see section \ref{subsec:cooling_model}). 
    The ratio between the dust surface density $\Sigma_{\rm d}$ and $\Sigma_{\rm g}$ is fixed to the interstellar dust abundance of $1\%$, whereas the local dust-to-gas ratio is allowed to vary with $z$ considering dust settling.
 
    We consider two grain size-distribution models. 
    The first model is the single-sized model where all dust particles are assumed to have equal size $a_1$. 
    Formally, the size distribution for the single-sized model can be written as 
    \begin{equation} \label{eq:Sigmad_1}
        \frac{dN_{\rm d}(a)}{da} = \frac{3\Sigma_{\rm d}}{4\pi\rho_{\rm int}a_1^3} \delta(a-a_1)
    \end{equation}
    where $dN_{\rm d}(a)/da$ is the number surface density per unit particle size $a$, $\Sigma_{\rm d}$ is the total dust mass surface density, $\rho_{\rm int}$ is the grains' internal density, and $\delta$ is the delta function.
    Equation~\eqref{eq:Sigmad_1} satisfies the normalization
    \begin{equation}\label{eq:Sigmad}
        \Sigma_{\rm d} = \int m_{\rm d}\frac{dN_{\rm d}(a)}{da} da,
    \end{equation}    
    where $m_{\rm d} = (4\pi/3)\rho_{\rm int}a^3$ is the particle mass.
    The second model is the power-law model where the grain size distribution is given by
    \begin{equation}\label{eq:Sigmad_a}
        \frac{dN_{\rm d}(a)}{da} =
        \left\{\begin{array}{ll}
        {\displaystyle \frac{(12+3p)\Sigma_{\rm d}}{4\pi \rho_{\rm int}\left(a_{\rm max}^{4+p}-a_{\rm min}^{4+p}\right)}a^{p},} & a_{\rm min} < a < a_{\rm max}, \\
            0, & {\rm otherwise}, 
        \end{array}\right. 
    \end{equation}
    where $p~(\neq -4)$ is the slope of size distribution, and $\subsc{a}{min}$ and $\subsc{a}{max}$ are the minimum and maximum particle sizes, respectively.
    We fix $p = -3.5$ and $\subsc{a}{min} = 1~\micron$.
    The maximum particle size serves as a free parameter in this study (see also section \ref{subsec:computational_domain and_parameter_choices}).
    Equation~\eqref{eq:Sigmad_a} also fulfills equation~\eqref{eq:Sigmad}.

    Dust particles settle toward the midplane owing to stellar gravity and diffuse away from the midplane owing to turbulence.
    We assume that the turbulent diffusion coefficient is constant in the vertical direction.
    This approach is valid if the VSI is the dominant source of disk turbulence and determines dust vertical diffusion.
    Even if there is the linear-VSI stable layer near the midplane, VSI-driven turbulence can penetrate the stable midplane layer and thereby produce nearly constant turbulent vertical intensity in the vertical direction \citep{PfeilKlahr:2021nr,FukuharaOkuzumi+:2023aa}.
    This situation can be realized when the stable midplane layer thickness is thinner than two gas scale heights, or the thickness of the unstable layer above the stable midplane layer is thicker than a few gas scale heights \citep{FukuharaOkuzumi+:2023aa}.
    In this study, the results for all parameter ranges satisfy these conditions.
    We note that, however, the diffusion coefficient may not be uniform in the vertical direction when these conditions are broken.
    
    Assuming the balance between settling and diffusion, the vertical distribution of the particles can be written as 
    \citep{TakeuchiLin2002}
    \begin{equation}\label{eq:rhoda}
        \frac{dn_{\rm d}(a,z)}{da} = C_d(a) \exp{\left[-\frac{z^2}{2H_{\rm g}^2}-\frac{{\rm St}_{\rm mid}(a)}{\alpha_z}\paren{\exp\frac{z^2}{2H_{\rm g}^2}-1} \right]},
    \end{equation}
    where $dn_{\rm d}(a,z)/da$ is the particle number density per unit radius at height $z$, $\mathrm{St}_{{\rm mid}}(a)$ is the Stokes number of the particles at the midplane, $\alpha_z$ is a dimensionless parameter that characterizes the level of the dust vertical diffusion caused by turbulence, and $C_d(a)$ is the normalized constant determined by the condition $dN_{\rm d}(a)/da = \int (dn_{\rm d}(a,z)/da)dz$.
    Because most dust particles lie at the region of $z \ll H_{\rm g}$, the exponential factor in equation \eqref{eq:rhoda} can be approximated as $\exp[-z^2/(2H_{\rm d}^2)]$, yielding \citep{FukuharaOkuzumi+:2021ca}
    \begin{equation}\label{eq:C_d2}
       C_d(a) = \frac{1}{\sqrt{2\pi}H_{\rm d}(a,\alpha_z)}\frac{dN_{\rm d}(a)}{da}.
    \end{equation}
    Here, $H_{\rm d}(a,\alpha_z)$ is the scale height of particles with size $a$ given by \citep{Dubrulle+1995,YoudinLithwick2007}
    \begin{equation}\label{eq:Hd}
        H_{\rm d}(a,\alpha_z) = \left[1+\frac{\mathrm{St}_{{\rm mid}}(a)}{\alpha_z} \right]^{-1/2}H_{\rm g}.
    \end{equation}
    The Stokes number is the product of the stopping time and Keplerian frequency. Assuming that the particle radius is smaller than the mean free path of the disk gas molecules, gas drag onto the particles follows Epstein's law, which gives \citep[see, e.g., ][]{BirnstielDullemond+:2010ny}
    \begin{equation}\label{eq:St}
        \mathrm{St}_{{\rm mid}}(a) = \frac{\pi \subsc{\rho}{int} a}{2\Sigma_{\rm g}}.
    \end{equation}

    The vertical--size distribution $dn_{\rm d}(a,z)/da$ gives the collisional heat transfer and opacity, which control the cooling timescale, as a function of $z$. The mean travel length of gas molecules colliding with dust particles $\subsc{\ell}{gd}$ is given by 
    \begin{equation}\label{eq:l_dg}
        \subsc{\ell}{gd} = \left(\int\pi a^2\frac{dn_{\rm d}}{da}  da\right)^{-1}.
    \end{equation}
    This determines the timescale of collisional heat transfer (see section \ref{subsec:cooling_model}).

    To calculate the cooling time, we use the Plank mean opacity and Rosseland mean opacity defined by
    \begin{equation}
        \kappa_{\rm P} = \frac{\displaystyle \int_0^\infty \kappa_{{\rm g},\lambda} B_\lambda (T) d\lambda}{\displaystyle \int_0^\infty B_\lambda (T) d\lambda},
    \end{equation}
    and
    \begin{equation}
        \frac{1}{\kappa_{\rm R}} = \frac{\displaystyle \int_0^\infty \frac{1}{\kappa_{{\rm g},\lambda}} \frac{\pd B_\lambda (T)}{\pd T} d\lambda}{\displaystyle \int_0^\infty \frac{\pd B_\lambda (T)}{\pd T} d\lambda},
    \end{equation}
    respectively, where $\kappa_{{\rm g},\lambda}$ is the wavelength-dependent opacity per gas mass, $\lambda$ is the wavelength, and $B_\lambda (T)$ is the Planck function.
    
    The opacity per gas mass is related to the size distribution of the dust particles as (e.g., \citealt{KondoOkuzumi+:2023aa})
    \begin{equation}
        \kappa_{{\rm g},\lambda} = \frac{1}{\rho_{\rm g}} \int \kappa_{{\rm d},\lambda}(a)m_{\rm d}\frac{dn_{\rm d}}{da} da,
    \end{equation}
    where $\kappa_{{\rm d},\lambda}(a)$ is the opacity per dust mass.
    Assuming that a dust particle is an uniform sphere, $\kappa_{{\rm d},\lambda}(a)$ can be written as \citep{KataokaOkuzumi+:2014js}
    \begin{equation}
        \kappa_{{\rm d},\lambda}(a) = \frac{\pi a^2}{m_{\rm d}}\times
        \left\{\begin{array}{ll}
        {\displaystyle \frac{24nkx}{\left(n^2+2 \right)^2},} & x \leq 1, \\
        {\displaystyle \min \left\{\frac{8kx}{3n}\left(n^3-\left(n^2-1 \right)^{3/2} \right),0.9 \right\},} & x > 1, 
        \end{array}\right. 
    \end{equation}
    where $x=2\pi a/\lambda$ is the size parameter and $n$ and $k$ are the real and imaginary parts of the complex refractive index, respectively.
    We calculate $n$ and $k$ at each wavelength using the Bruggeman mixing rule, assuming that the dust is a mixture of silicate and ice with a mass mixing ratio of 1:1 with no porosity.
    The values of $n$ and $k$ for silicate and ice are taken from \citet{Draine:2003xf} and \citet{WarrenBrandt:2008aa}, respectively.
    The internal density of the dust grains is then $\rho_{\rm int}=1.46 ~{\rm g~cm^{-3}}$.

    \subsection{Cooling model}\label{subsec:cooling_model}
    The spatial profile of the gas disk cooling rate depends on the size and spatial distribution of the dust particles \citep{Malygin+2017,BarrancoPei+:2018kc,FukuharaOkuzumi+:2021ca}.
    Using the dust model described in section \ref{subsec:dust_model}, we approximate the local cooling (thermal relaxation) timescale $\tau_{\rm relax}$ as \citep{Malygin+2017,PfeilKlahr2019}
    \begin{equation}\label{eq:taurelax}
        \tau_{\rm relax}(z) = \max{\{\tau_{\rm diff}(z),~\tau_{\rm coll}(z),~\tau_{\rm emit}(z)\}},
    \end{equation}
    where $\tau_{\rm diff}$, $\tau_{\rm coll}$, and $\tau_{\rm emit}$ are the timescales of radiative diffusion, collisional heat transfer, and radiative cooling, respectively.

    The cooling time in the optically thick regime is dominated by the radiative diffusion timescale $\tau_{\rm diff}$ as \citep{Malygin+2017}
    \begin{equation}\label{eq:taudiff}
        \tau_{\rm diff} = \frac{1}{\bar{D}k^2},
    \end{equation}
    where $\bar{D}$ and $k$ are the effective energy diffusion coefficient and the wavenumber of the perturbation, respectively.
    The effective energy diffusion coefficient is given by
    \begin{equation}
        \bar{D} = \frac{\lambda_{\rm r}c}{\kappa_{\rm R}(T) \rho_{\rm g}}\frac{4\eta}{1+3\eta}
    \end{equation}
    where $\lambda_{\rm r}$ is the flux limiter, $c$ is the light speed, and $\eta$ is the ratio between radiation energy density $E_{\rm r}$ to combined radiation and internal energy density $E_{\rm int}$.
    Here, $\eta$ is given by $\eta = E_{\rm r}/(E_{\rm r} + E_{\rm int})$ with $E_{\rm r} = 4\sigma_{\rm SB} T^4/c$ and $E_{\rm int} = \rho_{\rm g} C_V T$, where $\subsc{\sigma}{SB}$ is the Stefan--Boltzmann constant and $C_V = 5k_{\rm B}/2m_{\rm g}$ is the specific heat at constant volume.
    The flux limiter is fixed to $\lambda_{\rm r} = 1/3$, which is the value for the optically thick limit \citep{LevermorePomraning1981}.
    For $k$, we assume that the thermal perturbation is determined by the wavenumber of the VSI-driven turbulence structure.
    The VSI unstable modes emerge when a radial perturbation wavenumber is larger than a vertical one (e.g., \citealt{ArltUrpin2004}), and typically takes $\sim 20/H_{\rm g}$ by previous simulations of VSI-driven turbulence\footnote{ Note that the radial wavenumber of the VSI-driven turbulence structure can depend on the density and opacity. The previous hydrodynamical simulations including radiative transport by \citet{StollKley2014} have shown that an increase in the density leads to a smaller radial wavelength of turbulence structure.} (see appendix of \citealt{PfeilKlahr:2021nr}).
    Therefore, we set $k=20/H_{\rm g}$.

    The timescale of collisional heat transfer is given by \citep{FukuharaOkuzumi+:2021ca}
    \begin{equation}\label{eq:taucoll}
        \subsc{\tau}{coll} = \frac{\subsc{\ell}{gd}}{ \subsc{v}{th}},
    \end{equation}
    where $\subsc{v}{th}$ is the mean relative velocity between the gas molecules and dust particles. 
    The relative velocity $\subsc{v}{th}$ can be approximated as the mean thermal speed of the molecules, 
    \begin{equation}\label{eq:vth}
        \subsc{v}{th} = \sqrt{\frac{8\kB T}{\pi m_{\rm g}}}.
    \end{equation}
    We note that this collisional timescale is an approximation to the actual thermal accommodation timescale of the gas molecules and dust particles, which differs by a factor of $\gamma/(\gamma-1)$ \citep{BurkeHollenbach:1983aa,BarrancoPei+:2018kc,PfeilBirnstiel+:2023aa}, where $\gamma$ is the heat capacity ratio.
    
    The radiative cooling timescale $\subsc{\tau}{emit}$ in the optically thin limit is given by \citep{Malygin+2017}
    \begin{equation}\label{eq:tauemit}
        \subsc{\tau}{emit} = \frac{C_V}{16\kappa_{\rm P}(T)\subsc{\sigma}{SB} T^3}.
    \end{equation}
    Both $\subsc{\ell}{gd}$ and $\kappa_{\rm P}$ depend on the local size distribution of the dust particles (see section \ref{subsec:dust_model}). 

    \subsection{Defining the VSI-unstable layer}\label{subsec:unstable_region}
    The cooling timescale determines the VSI-unstable layer because rapid cooling reduces buoyancy preventing instability growth (e.g., \citealt{NelsonGresselUmurhan2013}). 
    Following \cite{LinYoudin2015}, the criterion for the VSI can be expressed as
    \begin{equation}\label{eq:cooling_criterion}
        \tau_{\rm relax}(z) \lesssim \tau_{\rm crit}.
    \end{equation}
    Here, $\tau_{\rm crit}$ is the vertically global critical cooling timescale defined by
    \begin{equation}
        \tau_{\rm crit} = \frac{H_{\rm g}}{R}\frac{|q|}{\gamma-1}\Omega_{\rm K}^{-1},
    \end{equation}
    with $\gamma = 1.4$.
    By applying the VSI criterion of equation \eqref{eq:cooling_criterion} to each point in the disk, we search for the VSI-unstable and -stable layers.
    We apply this vertically global criterion instead of the local criterion [equation (4) in \citealt{LinYoudin2015}].
    This is because the intensity of VSI-driven turbulence is tightly correlated with the thicknesses of the VSI-unstable and -stable layers predicted by the global criterion (\citealt{FukuharaOkuzumi+:2023aa}, see also section \ref{subsec:dust_vertical_diffusion}).
    The unstable and stable layers exist on the upper ($z>0$) and lower ($z<0$) halves of the disk, symmetrically concerning the midplane ($z=0$, see the middle panel of figure \ref{fig:fig1_Schematic_view}).
    For the upper halves, we determine the height of the unstable layer's upper boundary $z_{\rm u}$ and the height of the midplane stable layer's upper boundary $z_{\rm s}$.
    When the midplane stable layer is absent, we set $z_{\rm s} = 0$.
    Following \citet{FukuharaOkuzumi+:2023aa}, we also define the thicknesses of the linearly stable and unstable layers as $\Delta L_{\rm s} = 2z_{\rm s}$ and $\Delta L_{\rm u} = 2z_{\rm u}-\Delta L_{\rm s}$, respectively (see figure \ref{fig:fig1_Schematic_view}).

    \subsection{Estimating dust vertical diffusion coefficient}\label{subsec:dust_vertical_diffusion}
    The thicknesses of the VSI-unstable and -stable layers can be used to determine the vertical diffusion of dust particles \citep{FukuharaOkuzumi+:2023aa}.
    VSI-driven turbulence generally exhibits the vertical gas motion that is uniform in the vertical direction (e.g., \citealt{PfeilKlahr:2021nr,FukuharaOkuzumi+:2023aa}).
    Therefore, we estimate the dimensionless dust vertical diffusion coefficient of VSI-driven turbulence $\alpha_{z,\rm VSI}$ as
    \begin{equation}\label{eq:alpha_z}
        \alpha_{z,\rm VSI} = \frac{\langle v_z^2 \rangle}{c_{\rm s}^2}\cdot \tau_{\rm corr}\Omega_{\rm K},
    \end{equation}
    where $\langle v_z^2 \rangle$ is the time-averaged squared gas vertical velocity at the midplane and $\tau_{\rm corr}$ is the correlation time of VSI-driven turbulence.
    The correlation time of VSI-driven turbulence is still indeterminate, ranging from $\tau_{\rm corr}\Omega_{\rm K} \sim 0.2$ to $\sim 20$ (\citealt{StollKley:2016vp,Flock:2020aa}, see also section 4.1 of \citealt{FukuharaOkuzumi+:2023aa}).
    In this study, we fix $\tau_{\rm corr}\Omega_{\rm K} = 1.0$.

    \begin{figure}[t]
        \begin{center}
        \includegraphics[width=\hsize,bb = 0 0 391 278]{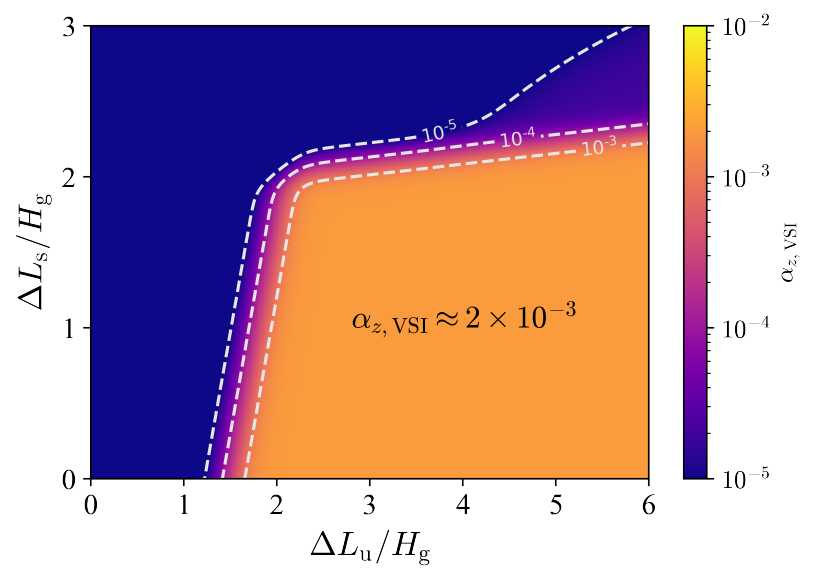}
        \end{center}
        \caption{Dimensionless dust vertical diffusion coefficient of VSI-driven turbulence $\alpha_{z,\rm VSI}$ [equation \eqref{eq:alpha_z}] as a function of $\Delta L_{\rm u}$ and $\Delta L_{\rm s}$. The dashed lines show $\alpha_{z,\rm VSI} = 10^{-3}$, $10^{-4}$, and $10^{-5}$.}
        \label{fig:fig2_alpha_z_VSI}
    \end{figure}

    Global axisymmetric simulations by \citet{FukuharaOkuzumi+:2023aa} show that the mean squared gas vertical velocity $\langle v_z^2 \rangle$ in VSI-driven turbulence is tightly correlated with the thicknesses of the VSI-unstable and -stable layers.
    We estimate $\langle v_z^2 \rangle$ from $\Delta L_{\rm u}$ and $\Delta L_{\rm s}$, using an empirical formula provided by \citet{FukuharaOkuzumi+:2023aa},
    \begin{equation}\label{eq:fitting_vz2}
        \frac{\langle v_z^2 \rangle}{c_{\rm s}^2}= f_{\rm T}(\Delta L_{\rm u},~\Delta L_{\rm s}) + f_{\rm pT}(\Delta L_{\rm u},~\Delta L_{\rm s}),
    \end{equation}
    where $f_{\rm T}$ and $f_{\rm pT}$ represent the dependence of $\Delta L_{\rm u}$ and $\Delta L_{\rm s}$ on $\langle v_z^2 \rangle$.
    For $f_{\rm T}$ and $f_{\rm pT}$, we use equations (17) and (19) of \citet{FukuharaOkuzumi+:2023aa}, which represent the sharp decrease of $\langle v_z^2 \rangle$ in $\Delta L_{\rm u} \lesssim 2H_{\rm g}$ and $\Delta L_{\rm s} \gtrsim 2H_{\rm g}$.
    In figure \ref{fig:fig2_alpha_z_VSI}, We plot $\alpha_{z,\rm VSI}$ [equation \eqref{eq:alpha_z}] as a function of $\Delta L_{\rm u}$ and $\Delta L_{\rm s}$.
    When $\Delta L_{\rm u}\gtrsim 2H_{\rm g}$ and $\Delta L_{\rm s} \lesssim 2H_{\rm g}$, equation \eqref{eq:alpha_z} predicts $\alpha_{z,\rm VSI} \approx 2\times 10^{-3}$; otherwise, $\alpha_{z,\rm VSI}$ approaches zero.

    \subsection{Calculation procedure}\label{subsec:Calculation_procedure}
    The vertical dust distribution depends on the turbulent diffusion coefficient $\alpha_z$ [equation~\eqref{eq:rhoda}]. 
    On the other hand, if we assume that the VSI is the main driver of disk turbulence, its diffusion strength $\alpha_{z,\rm VSI}$ depends on the disk's cooling structure [equation~\eqref{eq:alpha_z}] and in turn on vertical dust distribution. 
    In general, $\alpha_{z,\rm VSI}$ produced by VSI-driven turbulence under a given vertical dust distribution does not necessarily match $\alpha_z$ required to maintain the dust distribution. 
    We search for an equilibrium state where the two turbulence strengths match in the following steps.
    
    \begin{enumerate}
        \item For a given grain size distribution and radial position $R$, use equation~\eqref{eq:rhoda} to generate vertical dust profiles for various values of trial turbulence strengths $\alpha_{z} = \alpha_{z,\rm assume}$. The trial values are generated by dividing the range $10^{-8}\leq \alpha_{z,\rm assume}\leq 10^{-2}$ into grids of an equal logarithmic interval of 0.05.
        \item For each value of $\alpha_{z,\rm assume}$, use the cooling model [equation~\eqref{eq:taurelax}], VSI criterion [equation~\eqref{eq:cooling_criterion}], and the empirical formula for VSI-driven turbulence strength [equation~\eqref{eq:alpha_z}] to evaluate $\alpha_{z,\rm VSI}$.
        \item Interpolate $\alpha_{z,\rm VSI}$ as a smooth function of $\alpha_{z,\rm assume}$, and search for self-consistent equilibrium solutions where the predicted VSI-driven turbulence strength $\alpha_{z,\rm VSI}$ equals the assume turbulence strength $\alpha_{z,\rm assume}$. 
        Below, we denote the vertical diffusion coefficient for an equilibrium solution by $\alpha_{z,\rm equi}$. If $\alpha_{z,\rm VSI}<\alpha_{z,\rm assume}$ for all values of $\alpha_{z,\rm assume}$, VSI-driven turbulence cannot be strong enough to sustain the assumed dust vertical distribution, meaning that it cannot stop grain settling. We call this runaway dust settling.
    \end{enumerate}
    For every set of grain size and $R$, we repeat this procedure and search for equilibrium solutions.

    We ignore that VSI-driven turbulence revives before the dust settles toward the midplane in the runaway fashion.
    This assumption is valid if the dust-settling timescale is longer than the growth timescale of turbulence.
    The settling timescale can be estimated as $\sim z/|v_z| \approx {\rm St}_{\rm mid}^{-1}\Omega_{\rm K}^{-1}$ (e.g., \citealt{Dubrulle+1995,YoudinLithwick2007}), where $v_z \approx -{\rm St}_{\rm mid}\Omega_{\rm K}z$ is the vertical velocity of dust particles for the terminal velocity approximation.
    The typical growth timescale of VSI-driven turbulence is $\sim 10^2\Omega_{\rm K}^{-1}$ (e.g., \citealt{NelsonGresselUmurhan2013}), which is longer than the settling timescale for large dust grains with ${\rm St}_{\rm mid} \gtrsim 10^{-2}$.
    However, for small grains with ${\rm St}_{\rm mid} \lesssim 10^{-2}$, the growth timescale is shorter than the settling timescale, suggesting re-development of VSI-driven turbulence.
    In this case, weak turbulence may operate before the dust settles.

    \subsection{Parameter choices}\label{subsec:computational_domain and_parameter_choices}
    Our model involves two main parameters: the size of dust grains $a_1$ for the single-sized model or the maximum size of dust grains $a_{\rm max}$ for the power-law size distribution model, and the distance from the central star $R$.
    We divide the parameter ranges  $1~\micron < a < 1{\rm ~cm}$ or $1~\micron < a_{\rm max} < 1{\rm ~cm}$ and $1{\rm ~au} < R < 100{\rm ~au}$ into grids of an equal logarithmic interval of 0.1.
    
\section{Results}\label{sec:results}

In this section, we use the model described in section \ref{sec:method} to search for equilibrium solutions where dust settling balances with VSI-driven turbulent diffusion.
In section \ref{subsec:equilibrium_dust_vertical_profile_for_single_size_model}, we describe the properties of the equilibrium solutions and their dependence on the dust grain size and radial distance for the single-sized model.
In section \ref{subsec:equilibrium_dust_vertical_profile_for_size_distirbution_model}, we show the equilibrium solutions for the power-law size distribution model.

    \subsection{Equilibrium solutions for the single-sized model}\label{subsec:equilibrium_dust_vertical_profile_for_single_size_model}

    \begin{figure}[t]
        \begin{center}
        \includegraphics[width=\hsize,bb = 0 0 411 602]{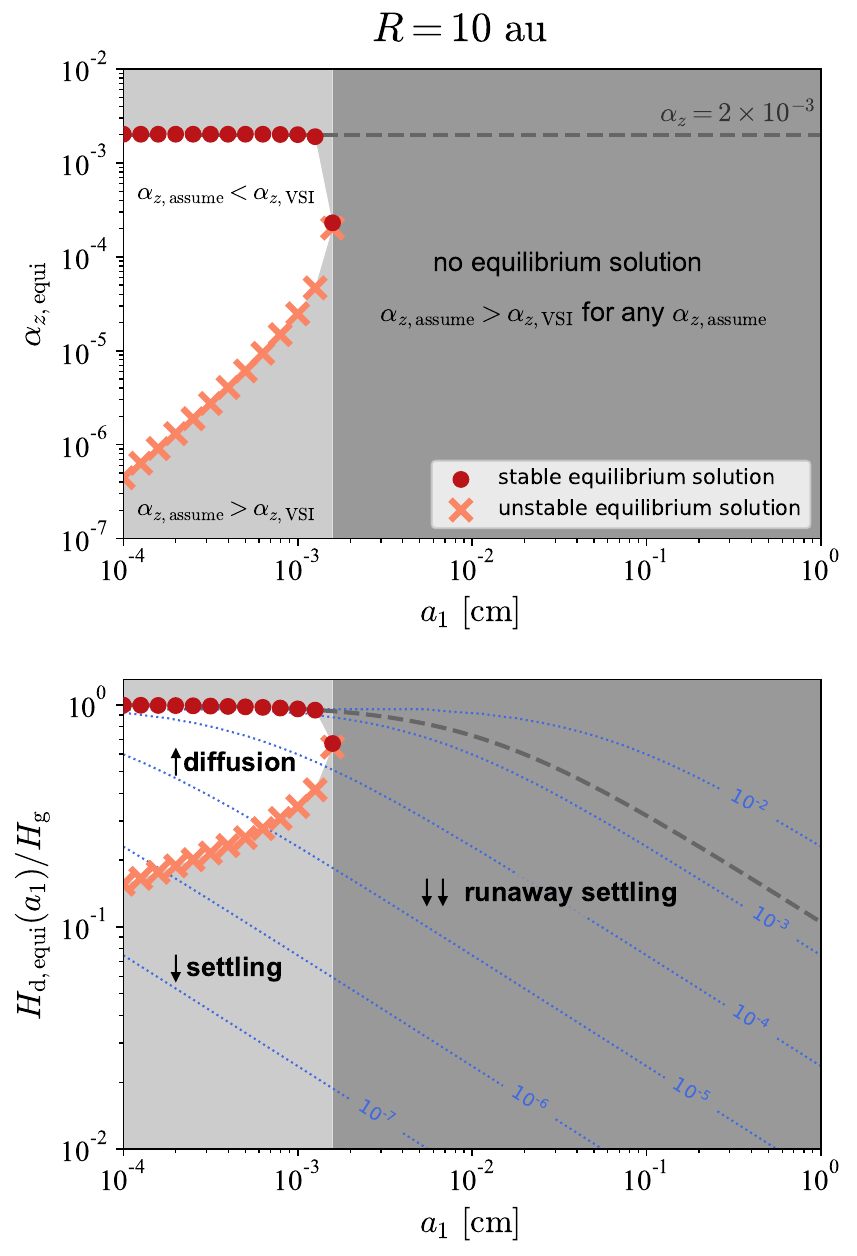}
        \end{center}
        \caption{Equilibrium vertical diffusion coefficient $\alpha_{z.\rm equi}$ (upper panel) and its dust scale height (lower panel), for different values of dust grain's size at $R=10~{\rm au}$ in the single-sized model. The symbols indicate stable (circles) and unstable (crosses) equilibrium solutions. The dashed lines show a constant value with $\alpha_{z} = 2\times 10^{-3}$. The dotted lines in the lower panel represent dust scale height with $\alpha_{z} = 10^{-2},~10^{-3},~10^{-4},~10^{-5},~10^{-6},$ and $10^{-7}$. }
        \label{fig:fig3_equi_a_Hdust_10au}
    \end{figure}

    First, we use the single-sized model at $R=10{\rm ~au}$ to illustrate the properties of the equilibrium solutions.
    Following the procedure described in section~\ref{subsec:Calculation_procedure}, we search for the equilibrium vertical diffusion coefficient $\alpha_{z,\rm equi}$ for different values of single grain size $a_1$.
    The upper panel of figure \ref{fig:fig3_equi_a_Hdust_10au} shows $\alpha_{z,\rm equi}$ for each $a_1$.
    We find that when $a_1\lesssim 10{\rm ~\micron}$, there exist two equilibrium solutions.
    The equilibrium solutions vanish when the grain size exceeds $\sim 10{\rm ~\micron}$.
    We also plot in the lower panel of figure \ref{fig:fig3_equi_a_Hdust_10au} the dust scale height with equilibrium solutions defined by $H_{\rm d,equi}(a) \equiv H_{\rm d}(a,\alpha_{z,\rm equi})$ as a function of $a_1$.

\begin{figure}[t]
        \begin{center}
        \includegraphics[width=\hsize,bb = 0 0 447 704]{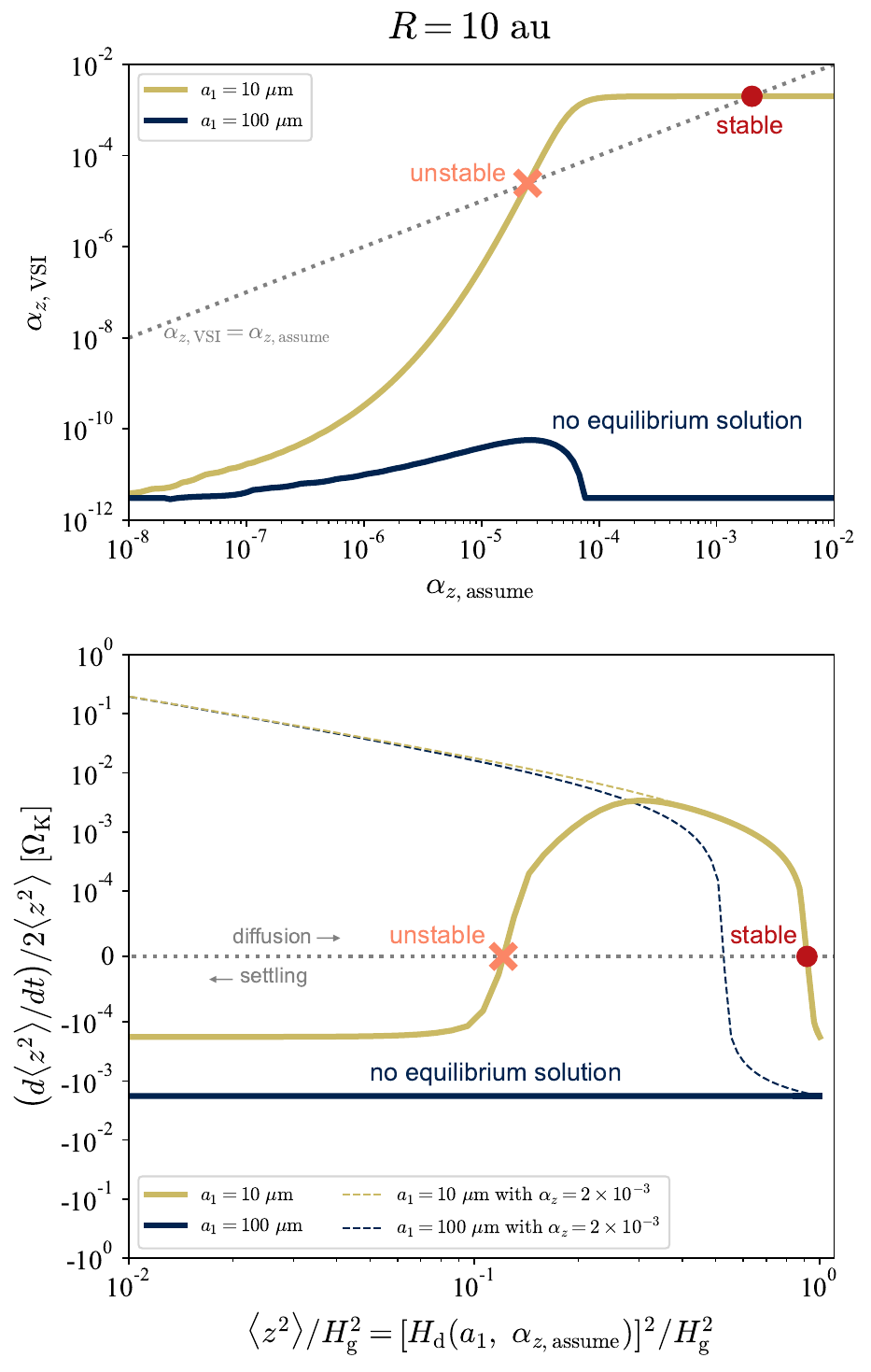}
        \end{center}
        \caption{Upper panel: VSI-driven vertical diffusion coefficient $\alpha_{z,\rm VSI}$ at $R = 10~\rm au$ as a function of $\alpha_{z,\rm assume}$ from the single-sized model with $a_1 = 10~\micron$ and $100~\micron$. The dotted line represents $\alpha_{z,\rm VSI}=\alpha_{z,\rm assume}$. The circle and cross symbols mark stable and unstable equilibrium solutions, respectively. Lower panel: $d\langle z^2\rangle/dt$ as a function of $\langle z^2\rangle$ for the two cases presented in the upper panel. The dashed lines show cases of a constant vertical diffusion coefficient as $\alpha_z = 2\times 10^{-3}$.}
        \label{fig:fig4_equilibrium_solution_10au}
    \end{figure}

    Figure \ref{fig:fig4_equilibrium_solution_10au} illustrates two cases at $R = 10~\rm au$ with  $a_1 = 10~{\rm \micron}$ and $100~{\rm \micron}$, representing the cases with two and no equilibrium solutions, respectively.
    The upper panel plots $\alpha_{z,\rm VSI}$ as a function of $\alpha_{z,\rm assume}$.
    For $a_1 = 10~{\rm \micron}$, $\alpha_{z,\rm VSI}$ drops sharply from $\approx 2\times 10^{-3}$ to $\ll 10^{-3}$ as $\alpha_{z,\rm assume}$ falls below $\sim 10^{-4}$.
    A decrease in $\alpha_{z,\rm assume}$ leads to dust depletion at high altitudes, yielding a decrease in the thickness of the VSI-unstable layer.
    This decrease in the unstable layer's thickness results in a strong suppression of VSI-driven turbulence around $\Delta L_{\rm u} \approx 2H_{\rm g}$ (see figure \ref{fig:fig2_alpha_z_VSI}).
    For $a_1 = 100~{\rm \micron}$, $\alpha_{z,\rm VSI}$ takes lower values of $\lesssim 10^{-10}$.
    This is because the VSI-unstable layer is thin ($\Delta L_{\rm u} \lesssim 2H_{\rm g}$) for any values of $\alpha_{z,\rm assume}$ due to dust vertical settling, resulting in suppression of VSI-driven turbulence.
    
    Equilibrium solutions correspond to the points where $\alpha_{z,\rm VSI} = \alpha_{z,\rm assume}$.
    For $a_1=10{\rm ~\micron}$, two equilibrium solutions exist at $\alpha_{z,\rm equi}\approx 2\times 10^{-3}$ and $2\times 10^{-5}$.
    For $a_1=100{\rm ~\micron}$, no equilibrium solution exists because $\alpha_{z,\rm VSI}<\alpha_{z,\rm assume}$ for all values of $\alpha_{z,\rm assume}$.
    This implies that the vertical diffusion by VSI-driven turbulence alone cannot sustain dust vertical distribution of any scale height in the latter case.

    To understand the results presented in the upper panel of figure \ref{fig:fig3_equi_a_Hdust_10au} in terms of the competition between vertical settling and turbulent diffusion, we consider a {\it time-dependent} toy model where the vertical distribution of dust grains evolves through settling and diffusion. 
    The equation that describes the evolution of the squared mean of the grains' vertical positions, $\langle z^2 \rangle$, is given by (for a derivation, see appendix \ref{appendix:vertical_diffusion_of_dust_particles})
    \begin{equation}\label{eq:g_z}
        \frac{1}{2}\frac{d\langle z^2\rangle}{dt} = -{\rm St}_{\rm mid}\Omega_{\rm K}\langle z^2\rangle + D_{z}(\alpha_{z,\rm VSI})\left(1-\frac{\langle z^2\rangle}{H_{\rm g}^2}\right),
    \end{equation}
    where $D_{z}(\alpha_z) = \alpha_{z}c_{\rm s}H_{\rm g}$ is the dust vertical diffusion coefficient.
    Below we simply call $\langle z^2\rangle^{1/2}$ the dust scale height. 
    In the right-hand side of equation \eqref{eq:g_z}, the first term represents settling toward the midplane at the terminal vertical velocity $-{\rm St}_{\rm mid}\Omega_{\rm K}z$. 
    The second term corresponds to the vertical diffusion by VSI-driven turbulence in a disk of gas scale height $H_{\rm g}$, with the factor $(1-{\langle z^2\rangle}/{H_{\rm g}^2})$ guaranteeing that the dust scale height never exceeds $H_{\rm g}$ in the limit of strong diffusion \citep{Ciesla:2010aa}.
    It can be shown from equation \eqref{eq:Hd} that our single-sized model assumes
    \begin{equation}
        \langle z^2 \rangle = \left[H_{\rm d}(a,\alpha_{z,\rm assume})\right]^2 = \left[1+\frac{{\rm St}_{\rm mid}(a)}{\alpha_{z,\rm assume}} \right]^{-1} H_{\rm g}^2
    \end{equation}
    for an arbitrary value of $\alpha_{z,\rm assume}$.
    Substituting this into equation~\eqref{eq:g_z}, we find that $d\langle z^2\rangle/dt=0$ if $\alpha_{z,\rm assume}=\alpha_{z,\rm VSI}$, meaning that the dust settling and diffusion indeed balance in the equilibrium solutions defined in section \ref{subsec:Calculation_procedure}. 
    This toy model also allows us to predict what would occur when $\alpha_{z,\rm assume}$ and $\alpha_{z,\rm VSI}$ are unequal. 
    For example, if $\alpha_{z,\rm assume} > \alpha_{z,\rm VSI}$ (i.e., if VSI-driven turbulence is not strong enough to sustain the vertical dust distribution), equation~\eqref{eq:g_z} shows that settling dominates over diffusion and the dust scale height decreases.

    We now use equation \eqref{eq:g_z} to explain why either two or no equilibrium solutions emerge depending on the value of $a_1$.
    The lower panel of figure \ref{fig:fig4_equilibrium_solution_10au} shows $d\langle z^2\rangle/dt$ as a function of $\langle z^2\rangle$ for $a_1 = 10$ and $100{\rm ~\micron}$.
    In general, there exists one equilibrium solution for a constant value of $D_z$ because $d\langle z^2\rangle/dt$ is then a monotonically decreasing function of $\langle z^2\rangle$, i.e., diffusion  (settling) tends to dominate at small (large) dust scale heights.
    This is illustrated by the dashed curves, which show $d\langle z^2\rangle/dt$ in the case where turbulence with $\alpha_{z,\rm VSI} \approx 2\times 10^{-3}$ is present independently of $\langle z^2\rangle$. 
    Therefore, multiple or no equilibrium solution occurs only when $D_z$ is a function of $\langle z^2\rangle$.  
    For $a_1 = 10{\rm ~\micron}$, the two equilibrium solutions have dust scale heights of $\langle z^2\rangle^{1/2} \sim H_{\rm g}$ and $0.3H_{\rm g}$. 
    Comparison with the dashed curve shows that the solution with higher $\langle z^2\rangle^{1/2}$ corresponds to the equilibrium solution for constant diffusion coefficient with $\alpha_{z,\rm VSI} \approx 2\times 10^{-3}$.     
    The solution with lower $\langle z^2\rangle^{1/2}$ emerges because VSI-driven turbulent diffusion is suppressed for small dust scale heights, i.e., for small $\alpha_{z, \rm assume}$, as mentioned earlier. 
    For $a_1 = 100~{\rm \micron}$, $d\langle z^2\rangle/dt$ is negative for all dust scale heights, indicating runaway dust settling.
    
    Equation \eqref{eq:g_z} can also be used to predict whether the equilibrium solutions are stable or not against a small perturbation in the dust scale height.
    In general, equilibrium solutions can be classified into stable and unstable ones.
    In this context, stable (unstable) solutions are the ones for which $d\langle z^2\rangle/dt$ becomes negative (positive) as we slightly increase $\langle z^2\rangle$ from the equilibrium value, so that the perturbation decays (diverges). 
    In other words, stable (unstable) solutions have a negative (positive) slope of $d\langle z^2\rangle/dt$ as a function of $\langle z^2\rangle$. 
    In the case of $a_1 = 10~\rm \mu m$, the equilibrium solution with the larger $\langle z^2\rangle$ is stable, whereas that with the smaller $\langle z^2\rangle$ is unstable (see figure~\ref{fig:fig4_equilibrium_solution_10au}).

    \begin{figure}[t]
        \begin{center}
        \includegraphics[width=\hsize,bb = 0 0 406 291]{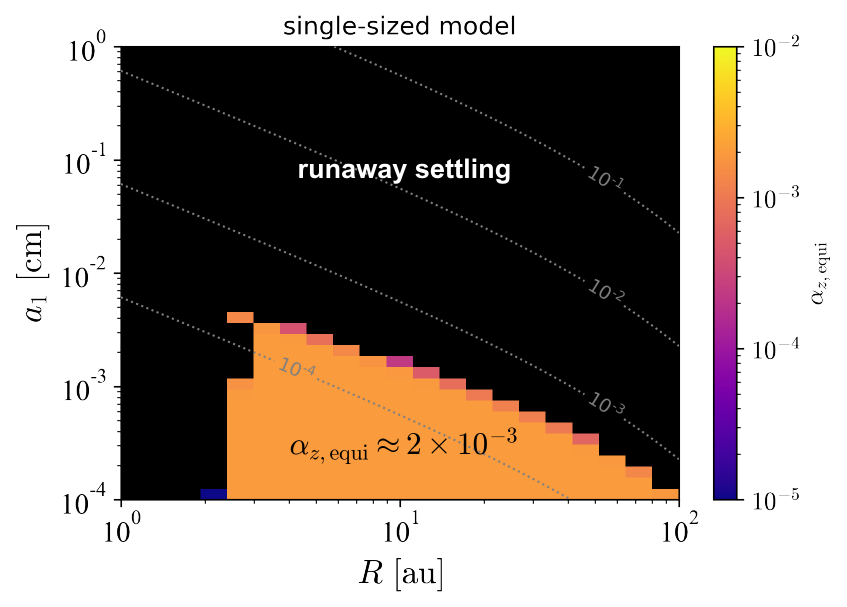}
        \end{center}
        \caption{Vertical diffusion coefficients $\alpha_{z,\rm equi}$ for the stable equilibrium solution from the single-sized model as a function of radial distance $R$ and grain size $a_1$. The black area indicates the parameter space where no equilibrium solution exists. In this area,  $\alpha_{z,\rm VSI}$ is lower than $\alpha_{z,\rm assume}$ for all $\alpha_{z,\rm assume}$, implying runaway dust settling. The dotted lines mark, from top to bottom, ${\rm St}_{\rm mid}(a_1) = 10^{-1},~10^{-2},~10^{-3},$ and $10^{-4}$.}
        \label{fig:fig5_equi_amax_alphaz_St_single}
    \end{figure} 

    \begin{figure}[t]
        \begin{center}
        \includegraphics[width=\hsize,bb = 0 0 395 290]{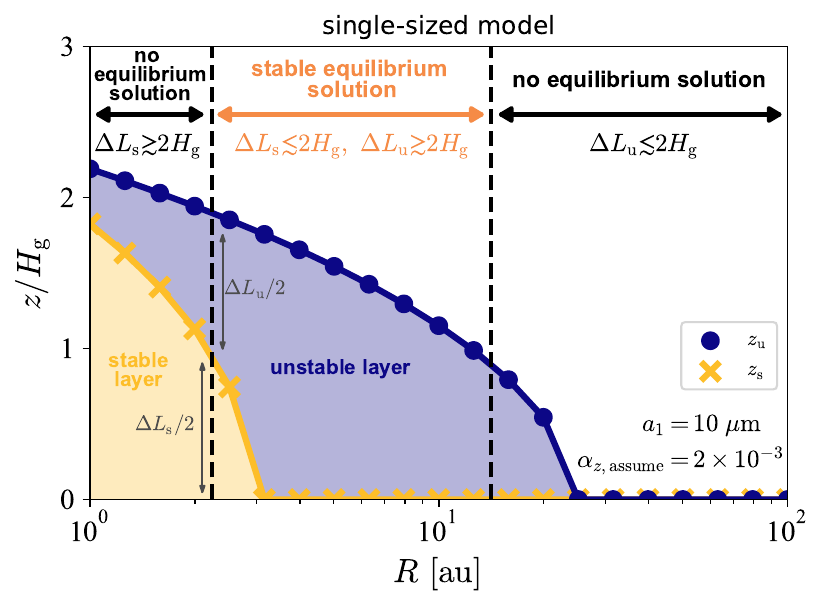}
        \end{center}
        \caption{Radial profiles of the height of the unstable layer's upper boundary (circles), $z_{\rm u}$, and the height of the midplane stable layer's upper boundary (crosses), $z_{\rm s}$, for $a_1 = 10{\rm ~\micron}$ with $\alpha_{z,\rm assume} = 2\times 10^{-3}$ from the single-sized model. The regions of $z_{\rm s} < z < z_{\rm u}$ and $0 < z < z_{\rm s}$ indicate the unstable and midplane stable layers, respectively. The left and right vertical dashed lines mark $\Delta L_{\rm s} (\equiv 2z_{\rm s}) \approx 2H_{\rm g}$ and $\Delta L_{\rm u} (\equiv 2z_{\rm u}-\Delta L_{\rm s}) \approx 2H_{\rm g}$, respectively.
        }
        \label{fig:fig6_Rz_LuLs_single_size}
    \end{figure}

    Next, we describe how the stable equilibrium solution varies with radial distance $R$. 
    Figure \ref{fig:fig5_equi_amax_alphaz_St_single} shows $\alpha_{z,\rm equi}$ for the stable solution as a function of $R$ and $a_1$ for the single-sized model. 
    The black area in the $R$--$a_1$ plane indicates the parameter space where no equilibrium solution exists and runaway dust settling is expected.
    The figure shows that equilibrium solutions only exist beyond $R \sim 3~\rm au$. 
    At $R \lesssim 3~\rm au$, the midplane region is significantly optically thick to its own thermal emission, yielding a thick VSI-stable layer ($\Delta L_{\rm s} \gtrsim 2H_{\rm g}$) around the midplane. 
    We map in figure \ref{fig:fig6_Rz_LuLs_single_size} the unstable and stable layers for $a_1 = 10{\rm ~\micron}$ with $\alpha_{z,\rm assume} = 2\times 10^{-3}$ from the single-sized model at the $R$--$z$ plane. 
    This figure indicates that the thickness of the stable midplane layer increases sharply as $R$ decreases from $\sim 3{\rm ~au}$.
    This thick stable layer prevents VSI-driven turbulence from developing ($\alpha_{z,\rm VSI}\ll 10^{-3}$; see figure \ref{fig:fig2_alpha_z_VSI}), resulting in runaway dust settling ($\alpha_{z,\rm VSI}<\alpha_{z,\rm assume}$ for all values of $\alpha_{z,\rm assume}$).  
    For the cases of large dust, optical depth decreases and the VSI-stable layer becomes thin.
    However, because the dust settles vertically, the VSI-unstable layer is also thin ($\Delta L_{\rm u} \lesssim 2H_{\rm g}$), leading to suppression of VSI-driven turbulence.
    The stable equilibrium solutions are generally accompanied by fully developed VSI-driven turbulence with $\alpha_{z,\rm VSI} \approx 2\times 10^{-3}$. 
    This is reasonable because any vulnerable turbulence whose strength decreases steeply with decreasing dust scale height would lead to an unstable solution.
    
    At $R \gtrsim 3~\rm au$, equilibrium solutions exist for sufficiently small $a_1$ as we already showed in the particular case of $R = 10~\rm au$. 
    The maximum grain size for equilibrium decreases with increasing $R$.
    This is because the dust number density decreases as $R$ increases: decreasing dust density leads to a longer cooling time, which makes the VSI-unstable layers vertically thinner (\citealt{Malygin+2017,PfeilKlahr2019,FukuharaOkuzumi+:2021ca}).

    \begin{figure}[t]
        \begin{center}
        \includegraphics[width=\hsize,bb = 0 0 450 290]{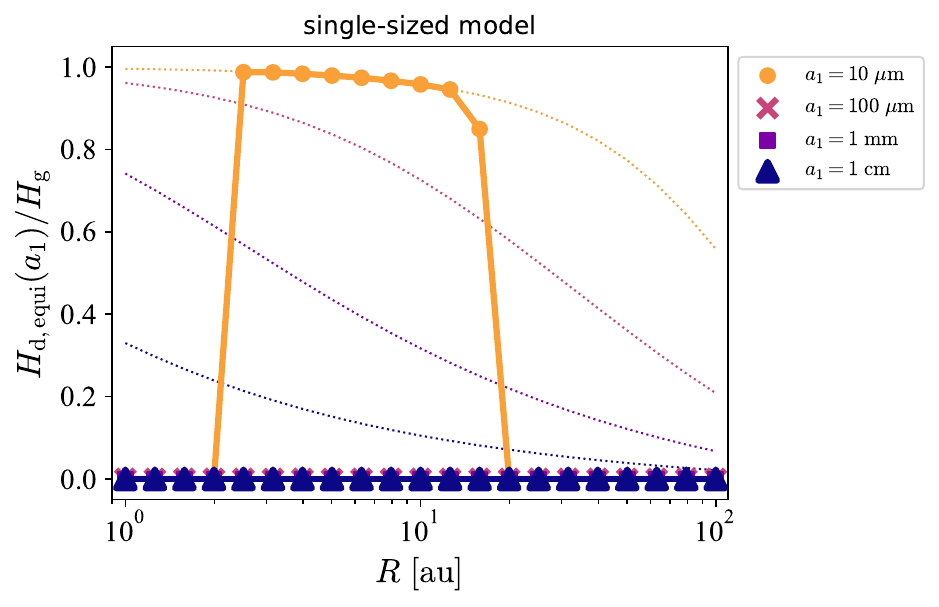}
        \end{center}
        \caption{Radial profiles of the dust scale height from the stable equilibrium solution for the single-sized dust model with different values of $a_1$. The dotted lines show the dust scale height for the fixed vertical diffusion coefficient of $\alpha_z = 2\times 10^{-3}$. 
        }
        \label{fig:fig7_Hdust_R_single}
    \end{figure}

    Figure \ref{fig:fig7_Hdust_R_single} shows the radial profiles of the dust scale height from the stable equilibrium solutions, $H_{\rm d,equi}(a_1)$, for different values of $a_1$.
    We set $H_{\rm d,equi}(a_1)$ to zero if runaway settling occurs.
    As mentioned above, the stable solutions are generally accompanied by fully developed VSI-driven turbulence with $\alpha_{z,\rm equi}\approx 2\times 10^{-3}$. 
    For reference, the dotted lines in the figure show the dust scale height for $\alpha_{z,\rm equi} = 2\times 10^{-3}$. 
    Because the stable solutions require small $a_1$, they always lead to a thick dust disk with $H_{\rm d,equi}(a_1) \approx H_{\rm g}$.

    Because the dust scale height is a function of ${\rm St}_{\rm mid}(a_1)$, we can expect that the value of ${\rm St}_{\rm mid}(a_1)$ rather than $a_1$ more directly determines whether there exist equilibrium solutions or not. 
    To confirm this expectation, we overplot in figure~\ref{fig:fig5_equi_amax_alphaz_St_single} contours of constant ${\rm St}_{\rm mid}(a_1)$. 
    We find that the equilibrium solution at $R \gtrsim 3~\rm au$ vanishes for ${\rm St}_{\rm mid}(a_1) \gtrsim 3\times 10^{-4}$. 

    \subsection{Equilibrium solutions for the power-law size distribution model}\label{subsec:equilibrium_dust_vertical_profile_for_size_distirbution_model}

    The grain size distribution affects the cooling timescale profile.
    Of the three timescales determining the cooling timescale (see section \ref{subsec:cooling_model}), the timescales of collisional heat transfer $\tau_{\rm coll}$ and radiative diffusion $\tau_{\rm diff}$ determine the thicknesses of the unstable and midplane stable layers, respectively.
    For the power-law size distribution with a slope of $p=-3.5$, the largest grains dominate the dust mass budget, whereas the smallest grains dominate the total geometric cross-section and thereby control collisional heat transfer.
    Therefore, even if large grains exist, small grains can still sustain the unstable layer.

    \begin{figure}[t]
        \begin{center}
        \includegraphics[width=\hsize,bb = 0 0 406 291]{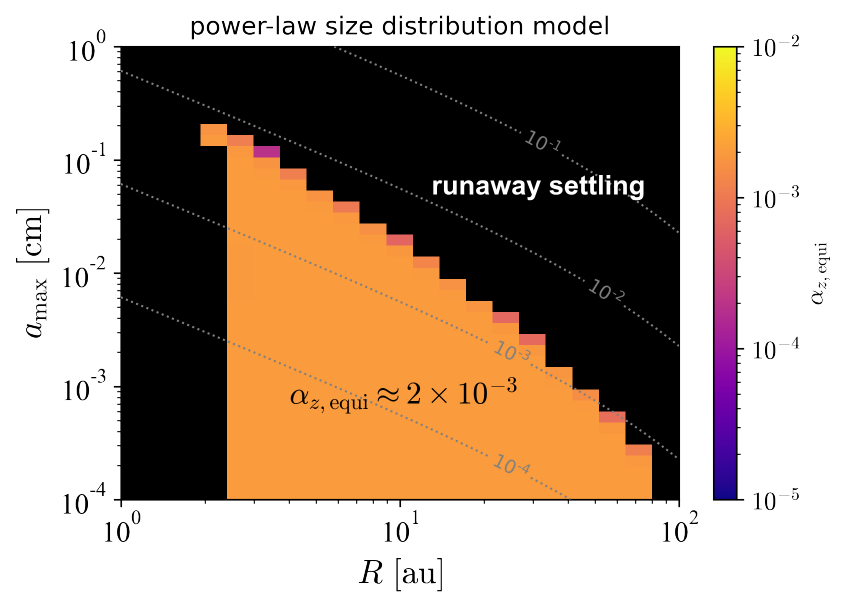}
        \end{center}
        \caption{Same as figure \ref{fig:fig5_equi_amax_alphaz_St_single}, but from the power-law size distribution model. The dotted lines mark, from top to bottom, ${\rm St}_{\rm mid}(a_{\rm max}) = 10^{-1},~10^{-2},~10^{-3},$ and $10^{-4}$.}
        \label{fig:fig8_equi_amax_alphaz_St_size}
    \end{figure}    

    Figure \ref{fig:fig8_equi_amax_alphaz_St_size} illustrates how the small grains in the power-law size distribution model extend the parameter space where strong VSI-driven turbulence is sustained. 
    This figure shows $\alpha_{z,\rm equi}$ for the stable equilibrium solutions as a function of $R$ and $a_{\rm max}$ for the power-law size distribution model.
    At $R = 10{\rm ~au}$, the equilibrium solutions exist up to $a_{\rm max} = 100~\rm \mu m$. 
    This is in contrast to the single-sized model, which only allows equilibrium solutions up to $10{\rm ~\micron}$ at the same radial position (see figure \ref{fig:fig5_equi_amax_alphaz_St_single}).
    In terms of the Stokes number, the power-law size distribution model allows the equilibrium solutions up to ${\rm St}_{\rm mid}(a_{\rm max}) \approx 10^{-3}$--$10^{-2}$ depending on $R$. 
    As in the single-sized model, the stable solution is accompanied by fully developed VSI-driven turbulence with $\alpha_{z,\rm equi} \approx 2\times 10^{-3}$.
    At $R\lesssim 3~{\rm au}$ or at sufficiently large $a_{\rm max}$, the equilibrium solution vanishes and runaway settling ($\alpha_{z,\rm assume} > \alpha_{z,\rm VSI}$) occurs for the same reason as in the single-sized model.

    \begin{figure}[t]
        \begin{center}
        \includegraphics[width=\hsize,bb = 0 0 460 290]{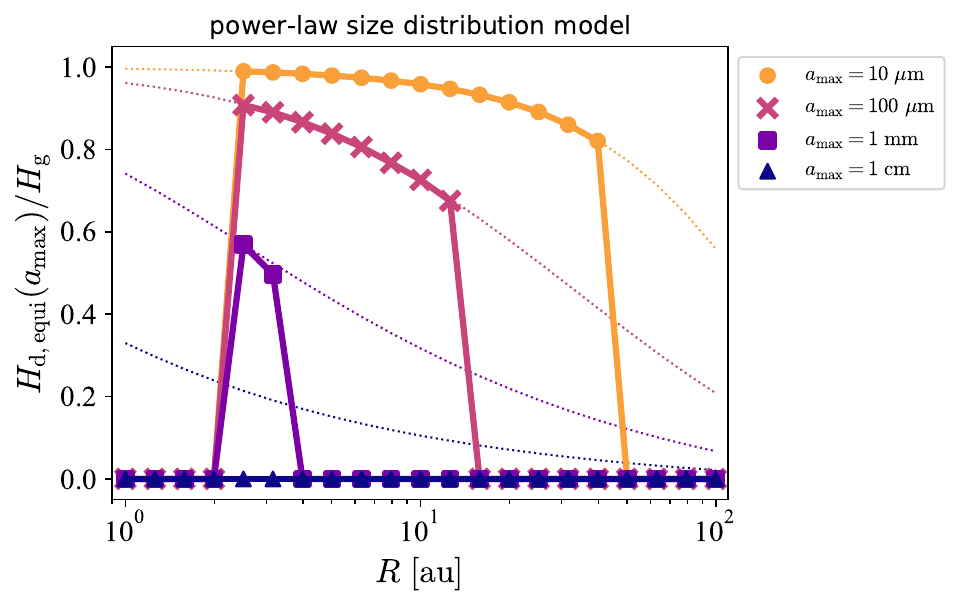}
        \end{center}
        \caption{Same as figure \ref{fig:fig7_Hdust_R_single}, but from the power-law size distribution model. 
        }
        \label{fig:fig9_Hdust_R_size}
    \end{figure}

    Figure \ref{fig:fig9_Hdust_R_size} plots the dust scale height for the stable equilibrium solution as a function of $R$ from the power-law size distribution model with different values of $a_{\rm max}$. 
    As in figure \ref{fig:fig7_Hdust_R_single}, we set $H_{\rm d, equi} = 0$ for runaway settling.
    For $a_{\rm max}=10~{\rm \micron}$, VSI-driven turbulence maintains a thick dust layer of $H_{\rm d,equi} \gtrsim 0.8H_{\rm g}$ at $2{\rm ~au}\lesssim R \lesssim 40{\rm ~au}$.
    As $a_{\rm max}$ increases, the region with the stable equilibrium solution shrinks, and the equilibrium dust scale height within that region decreases because the settling velocity increases while keeping $\alpha_{z,\rm equi} \approx 2\times 10^{-3}$.
    For $a_{\rm max} > 1~{\rm mm}$, the equilibrium solution vanishes at all $R$.
    
\section{Discussion}\label{sec:discussion}
    \subsection{Implications for disk observations}\label{subsec:implications_for_disk_observation}

    In section \ref{sec:results}, we have shown that the dust grain size and radial distance determine whether the VSI can sustain high dust diffusion.
    VSI-driven turbulence can maintain a thick dust layer of $H_{\rm d} \approx H_{\rm g}$ within an annular region with a moderate optical depth and small grains.
    However, when both the dust size and radial distance are large, VSI-driven turbulence is suppressed, resulting in runaway dust settling leading to $H_{\rm d} \approx 0$.
    
    These differences in dust settling levels may provide an explanation for varying degrees of dust settling as inferred from recent radio interferometric observations of some protoplanetary disks.
    The disks around HL Tau and Oph 163131 exhibit well-defined dust gaps, limiting the dust scale height to $\lesssim 0.1H_{\rm g}$ at $R \gtrsim 10{\rm ~au}$ \citep{Pinte:2016aa} and at $R\approx 100{\rm ~au}$ \citep{VillenaveStapelfeldt+:2022pp}, respectively.
    These estimated dust scale heights are in line with runaway dust settling in the outer disk regions predicted in this study.
    In contrast, the disk around HD 163296 has two major dust rings at $70$ and $100{\rm ~au}$ that show high and low degrees of dust settling with $H_{\rm d}/H_{\rm g} \gtrsim 0.8$ and $\lesssim 0.1$, respectively \citep{DoiKataoka:2021oz}.
    This difference in dust diffusion by radial distance is consistent with the trend in the radial profile of dust scale heights predicted by our model (see figures \ref{fig:fig7_Hdust_R_single} and \ref{fig:fig9_Hdust_R_size}).
    Therefore, we hypothesize that VSI-driven turbulence dominates vertical dust diffusion in these disks, and that this turbulence is suppressed in the outer disk regions.
    Testing this hypothesis requires detailed modeling of gas, dust, and cooling rate profiles in these disks.
    
    \subsection{Implications for disk and dust evolution}\label{subsec:implications_for_disk_and_dust_evolution}

    In section \ref{sec:results}, we have shown that there exists a parameter space where VSI-driven turbulence is too weak to stop dust settling. 
    This occurs either when the VSI-stable layer at the midplane is too thick or when the VSI-unstable layer is too thin for the VSI to operate. 
    The former condition is met at small orbital radii ($R \lesssim$ 3 au), whereas the latter condition is met when the grains grow beyond a certain size.
    
    This runaway settling is beneficial for dust growth and planetesimal formation.
    A high degree of dust settling generally promotes planetesimal formation through the streaming and gravitational instabilities (e.g., \citealt{Sekiya:1998aa,Youdin:2002aa,Johansen:2009aa,GoleSimon+:2020aa,UmurhanEstrada+:2020yi,ChenLin:2020kh}).
    Weak turbulence leading to high dust concentration is also conducive to dust growth through coagulation without collisional fragmentation and erosion (e.g., \citealt{Brauer:2008aa,Okuzumi:2012aa}).
    
    Figures \ref{fig:fig7_Hdust_R_single} and \ref{fig:fig9_Hdust_R_size} imply that strong dust diffusion by the VSI tends to occur in an annular region where the optical depth is moderate (see also figure \ref{fig:fig6_Rz_LuLs_single_size}). 
    The vertically extended dust in this annulus may block the radiation of the central star and thereby cast a shadow beyond the annulus \citep[e.g.,][]{DullemondDominik+:2001aa,DullemondDominik2004}. 
    The shadowing generally results in a significant drop in temperature and therefore can affect the chemical evolution in that region \citep{OhnoUeda:2021aa,NotsuOhno+:2022aa}.

    The inner and outer edges of the VSI-turbulence zone also serve as potential sites for planetesimal formation. 
    A steep radial change in turbulence viscosity at the edges may trigger the Rossby wave instability \citep{LovelaceLi+:1999sg,LiFinn+:2000aa,LiColgate+:2001gv} and create a long-lived vortex, promoting dust concentration and subsequent planetesimal formation through gravitational collapse (e.g., \citealp{BargeSommeria:1995qd}).
    The sharp drop in turbulent diffusivity at the inner edge of the VSI-driven turbulence zone can also trigger a runaway pile-up of dust grains \citep{HyodoIda+:2021aa,HyodoIda+:2022aa} because the increase of dust-to-gas ratio at the midplane reduces the radial drift velocity of dust.
    Moreover, if the turbulent viscosity caused by the VSI dominates gas disk accretion, the outer edge of VSI-driven turbulence may create a local maximum in the radial profile of the gas pressure because the turbulent viscosity decreases sharply toward the outside.
    This is similar to the mechanism by which a pressure maximum is generated near the dead-zone inner edge of MRI (e.g., \citealt{DzyurkevichFlock+:2010aa,FlockFromang+:2016aa,Flock:2017aa}).
    The pressure maximum can trap dust particles \citep{Whipple:1972vv,Adachi:1976uv,Weidenschilling:1977wt}.
    All these dust concentration processes could lead to planetesimal formation via the streaming and the gravitational instabilities (e.g., \citealt{YoudinGoodman:2005aa,JohansenYoudin2007,Johansen:2009aa,Carrera:2015aa,Yang:2017aa}).
    We plan to quantify the impact of these dust concentration processes on planetesimal formation by conducting hydrodynamical simulations around the edge of the VSI-driven turbulence zone.

    \subsection{Effects of minimum grain size, grain size distribution, and disk mass}\label{subsec:Effects_of_dust_size_distribution_and_disk_mass}
    
    The results presented in section \ref{subsec:equilibrium_dust_vertical_profile_for_size_distirbution_model} depend on the minimum grain size and slope of the size distribution because the small grains dominate the cooling rate.
    In this study, we have fixed the minimum size of grains and slope of grain size distribution to $a_{\rm min} = 1~{\rm \micron}$ and $p=-3.5$, respectively.
    Smaller dust grows quickly through Brownian motion \citep{BirnstielOrmel+:2011aa}, but its size limit is uncertain, approximately ranging from $0.1$ to $1 {\rm ~\micron}$.
    As $a_{\rm min}$ decreases, the cooling timescale decreases, leading to a vertically more extended VSI-unstable layer.
    The grain size distribution limited by the radial drift can also become significantly steeper (e.g., \citealt{BirnstielOrmel+:2011aa,StammlerBirnstiel:2022ox,Birnstiel:2023aa}), with $p$ reaching approximately $-2.5$.
    As $p$ increases, cooling would be less efficient because the number of the smallest grains is smaller, leading to more suppressed the VSI.
    
    Furthermore, variations in the disk mass can alter the cooling rate profile.
    In this study, we have fixed the disk mass to $M_{\rm disk}=0.01M_\odot$.
    The disk mass can depend on disk age \citep{CazzolettiManara+:2019aa,TestiNatta+:2022aa}; in particular, the disk mass of the young disk around HL Tau can be estimated as $M_{\rm disk}\sim 0.1M_\odot$ \citep{Kwon:2015aa}.
    The massive disk would extend the parameter space where the equilibrium solutions exist because the larger disk mass can make cooling more efficient, leading to the radial and vertical expansion of the VSI-unstable layers (see section 4.3 and figure 10 of \citealp{FukuharaOkuzumi+:2021ca}).
    However, as the disk mass increases, the region with no equilibrium solutions due to the thick stable layer around the midplane, corresponding to the regions with $R\lesssim 3 ~{\rm au}$ for figures \ref{fig:fig5_equi_amax_alphaz_St_single} and \ref{fig:fig8_equi_amax_alphaz_St_size}, would be more extended to larger radial distance.
    This is because as the dust density increases, optical depth increases, leading to inefficient cooling.
    
    \subsection{Limitations of the model}\label{subsec:limit_of_this_study}
    So far we have assumed the local model in the radial direction, meaning that gas and dust remain stationary radially.
    However, VSI-driven turbulence can diffuse dust radially, altering its spatial distribution \citep{StollKley:2016vp,Flock:2020aa,DullemondZiampras+:2022aa}.
    This effect implies that the edges of the VSI-driven turbulence zone evolve with dust.
    Quantifying this effect should be studied in hydrodynamical simulations that include the dynamic and thermal coupling between gas and dust.

    In the region where dust settles in a runaway fashion, other mechanisms that drive disk turbulence may also contribute to the vertical diffusion of dust grains.
    The MRI can drive gas turbulence in both the inner and outer regions of protoplanetary disks because of the high ionization of gas (for reviews, \citealt{TurnerFromang+:2014aa,LesurFlock+:2023aa}).
    The streaming instability caused by strong dust settling can also trigger turbulent gas motion \citep{JohansenYoudin2007,YangZhu:2021aa}.
    Turbulence driven by these mechanisms can maintain a vertical dust distribution that is balanced between turbulent diffusion and settling.
    Moreover, the dust vertical diffusion due to these other mechanisms may revive VSI-driven turbulence through changes in the vertical profiles of dust particles and the cooling rate.
    To understand this, we should investigate which mechanism dominates turbulence generation in each region of protoplanetary disks.
    
    Furthermore, this study ignores the effects of dust, magnetic field, and vertical thermal structure.
    Dust would increase the effective buoyancy frequency of the gas \citep{Lin:2017aa} that prevents the growth of the linear VSI.
    The gas--dust drag force can also suppress the VSI and dust vertical diffusion \citep{Lin:2019aa,LehmannLin:2022nr,LehmannLin:2023aa}.
    Moreover, magnetic fields threading the global disk may suppress the VSI either directly through magnetic tension or indirectly through MRI turbulence \citep{NelsonGresselUmurhan2013,LatterPapaloizou2018,Cui:2020aa}.
    The roles of magnetic fields in the VSI suppression can be positive or negative depending on non-ideal magnetohydrodynamical effects (ambipolar diffusion, Ohmic resistivity, and Hall effect; \citealt{Cui:2020aa,CuiBai:2022aa,CuiLin:2021cj,LatterKunz:2022ic}).
    Additionally, \citet{ZhangZhu+:2024aa} recently found that vertical thermal stratification with a colder interior and a hotter surface can suppress VSI-driven turbulence around the midplane.
    They may change the levels of the equilibrium vertical dust profile.

\section{Summary}\label{sec:summary}
We have searched for the equilibrium vertical dust profile where settling balances with diffusion caused by VSI-driven turbulence.
We construct the semi-analytic model that determines the vertical profile of dust grains and the intensity of VSI-driven turbulence in a self-consistent manner (figure \ref{fig:fig1_Schematic_view}).
Our key findings are summarized as follows.
\begin{enumerate}
    \item We find that there exist equilibrium solutions where dust settling balances with VSI-driven turbulent diffusion for small grains (figure \ref{fig:fig3_equi_a_Hdust_10au}). If we assume that all grains have equal size, there exist two equilibrium solutions when the single grain size is smaller than $10{\rm ~\micron}$ at $10{\rm ~au}$. If the grain size exceeds $\sim 10{\rm ~\micron}$, the equilibrium solutions vanish.
    \item For the cases of small grains, two equilibrium solutions are classified into stable and unstable ones [equation \eqref{eq:g_z} and figure \ref{fig:fig4_equilibrium_solution_10au}]. The stable ones correspond to the dust scale height of $\alpha_{\rm z,equi}\approx 2\times 10^{-3}$, where $\alpha_{\rm z,equi}$ is the equilibrium dimensionless vertical diffusion coefficient. For the cases of large grains, no equilibrium solutions indicate runaway dust settling.
    \item The existence of the equilibrium solutions depends on the radial distance $R$ as well as dust size (figure \ref{fig:fig5_equi_amax_alphaz_St_single}). The equilibrium solutions only exist beyond $R\sim 3{\rm ~au}$ because the midplane region at $R\lesssim 3{\rm ~au}$ is optically thick, yielding a thick VSI-stable layer (figure \ref{fig:fig6_Rz_LuLs_single_size}). The maximum grain size that allows for the equilibrium solutions also decreases with increasing $R$. Because the equilibrium solutions require a small grain size, they lead to a thick dust disk with $H_{\rm d,equi}\approx 0.8H_{\rm g}$ (figure \ref{fig:fig7_Hdust_R_single}), where $H_{\rm d,equi}$ is the dust scale height with the equilibrium solutions.
    \item If the particle size distribution is assumed to follow a power law, the small grains extend the parameter space where strong VSI-driven turbulence is sustained (figure \ref{fig:fig8_equi_amax_alphaz_St_size}). At $10~{\rm au}$, the equilibrium solutions exist up to $a_{\rm max} = 100{\rm ~\micron}$, where $a_{\rm max}$ is the maximum size of dust grains. For $a_{\rm max}=10~{\rm \micron}$ in the entire disk, VSI-driven turbulence maintains a thick dust layer of $H_{\rm d,equi}\gtrsim 0.8H_{\rm g}$ at $2{\rm ~au}\lesssim R \lesssim 50{\rm ~au}$ (figure \ref{fig:fig9_Hdust_R_size}).
\end{enumerate}

Our results suggest that dust diffusion by VSI-driven turbulence has different levels depending on the radial distance.
This variation may explain the different degrees of dust settling inferred from observations of some protoplanetary disks.
This implies that VSI-driven turbulence plays a dominant role in vertical dust diffusion within these disks.
Testing this hypothesis requires a more quantitative investigation of these disks' cooling rate structure.

\begin{ack}
    We thank Tomohiro Ono for discussions that motivated this project.
    We also thank Akimasa Kataoka, Kiyoaki Doi, Hidekazu Tanaka, Mario Flock, and Takahiro Ueda for the useful discussions of applications to observed protoplanetary disks.
    We appreciate the anonymous referee for comments that greatly helped improve the manuscript.
    This work was supported by JSPS KAKENHI Grant Numbers 
    JP20H01948, JP20H00182, JP22KJ1337, JP23H01227, and JP23K25923.
\end{ack}

\bibliographystyle{apj}
\bibliography{FukuharaOkuzumi24}

\appendix

\section{Vertical diffusion of dust particles}\label{appendix:vertical_diffusion_of_dust_particles}
The evolution equation of the squared mean of the grains’ vertical positions [equation \eqref{eq:g_z}] is derived by considering the flux of dust in the vertical direction.
The vertical diffusion of dust density $\rho_{\rm d}$ in a turbulent protoplanetary disk is given by \citep{Dubrulle+1995,FromangPapaloizou:2006rz}
\begin{equation}\label{eq:ap_vertical_motion}
    \frac{\pd \rho_{\rm d}}{\pd t} = - \frac{\pd}{\pd z}\left(\rho_{\rm d}v_z\right) + \frac{\pd}{\pd z}\left[\rho_{\rm g}D_z\frac{\pd}{\pd z}\left(\frac{\rho_{\rm d}}{\rho_{\rm g}}\right)\right],
\end{equation}
where $v_z$ is the vertical velocity of dust and $D_z$ is the vertical turbulent diffusion coefficient of dust.
Expanding the second term of the right side in equation \eqref{eq:ap_vertical_motion} and assuming vertical hydrostatic equilibrium for gas [equation \eqref{eq:rhogas}] yield \citep{Ciesla:2010aa}
\begin{equation}
    \frac{\pd \rho_{\rm d}}{\pd t} = - \frac{\pd}{\pd z}\left(\rho_{\rm d}v_z\right) + \frac{\pd}{\pd z}\left(D_z\frac{\pd \rho_{\rm d}}{\pd z}\right) + \frac{\pd}{\pd z}\left(D_z\frac{z}{H_{\rm g}^2}\rho_{\rm d}\right).
\end{equation}
Next, we multiply both sides of this equation by $z^2$ and integrate it from $z = -\infty$ to $z =\infty$.
The partial integrals are performed repeatedly and surface terms are dropped, resulting in
\begin{equation}\label{eq:ap_integration}
    \frac{1}{2}\frac{\pd}{\pd t}\int_{-\infty}^{\infty}\rho_{\rm d}z^2 dz = \int_{-\infty}^{\infty}\rho_{\rm d}v_zz dz + D_z\left( \int_{-\infty}^{\infty}\rho_{\rm d} dz - \frac{1}{H_{\rm g}^2}\int_{-\infty}^{\infty}\rho_{\rm d}z^2 dz \right).
\end{equation}
Assuming the terminal vertical velocity $v_z = -{\rm St}_{\rm mid}\Omega_{\rm K}z$ and a constant dust surface density defined by $\Sigma_{\rm d} \equiv \int \rho_{\rm d} dz$, equation \eqref{eq:ap_integration} can be written as 
\begin{equation}
    \frac{1}{2}\frac{d\langle z^2\rangle}{dt} = - {\rm St}_{\rm mid}\Omega_{\rm K}\langle z^2\rangle + D_z\left(1-\frac{\langle z^2\rangle}{H_{\rm g}^2}\right),
\end{equation}
where $\langle z^2 \rangle$ is the ensemble average of $z^2$ for dust particles defined by
\begin{equation}\label{eq:ap_z2}
    \langle z^2 \rangle \equiv \frac{\displaystyle\int_{-\infty}^{\infty}\rho_{\rm d}z^2 dz}{\displaystyle\int_{-\infty}^{\infty}\rho_{\rm d}dz}.
\end{equation}
If we assume the single-sized model and $\rho_{\rm d}\propto \exp{\left(-z^2/2H_{\rm d}^2\right)}$, equation \eqref{eq:ap_z2} results in $\langle z^2 \rangle^{1/2} = H_{\rm d}$, where $H_{\rm d}$ is the dust scale height.

We use the Stokes number at the midplane ${\rm St}_{\rm mid}$ to calculate the vertical velocity.
In reality, the Stokes number increases as $z$ increases, causing the small grains to settle toward the midplane faster.
However, in this study, the Stokes numbers of the smallest grains exceed unity only at $z \gtrsim 4H_{\rm g}$.
This height is higher than the height of the VSI-unstable layer ($z \lesssim 3H_{\rm g}$, see also figure \ref{fig:fig6_Rz_LuLs_single_size} in section \ref{subsec:equilibrium_dust_vertical_profile_for_single_size_model}).
Therefore, the effect of ignoring the regime for larger Stokes numbers than unity has little impact on our results.

\end{document}